\documentstyle[aps,prd]{revtex}
\begin{document}
\title{Towards a physical interpretation for the Stephani Universes.} 

\author{Roberto A. Sussman}
\address{Facultad de Ciencias, UAEM, Av.
Universidad 1001, Cuernavaca, Morelos, 62145, M\'exico.\footnote{On sabatical leave from
Instituto de Ciencias Nucleares,  Apartado Postal 70543, UNAM, M\'exico DF,
04510, M\'exico. Email sussman@nuclecu.unam.mx.}}   
\date{\today}

\maketitle   
                 
\begin{abstract}                 
A physicaly reasonable interpretation is provided for the perfect fluid,
sphericaly symmetric, conformally flat ``Stephani Universes''. The free
parameters of this class of exact solutions are determined so that the ideal gas
relation $p=nk_BT$ is identicaly fulfiled, while the full equation of state of 
a classical monatomic ideal gas and a matter-radiation mixture holds up to a
good approximation in a near dust, matter dominated regime. Only the models 
having spacelike slices with positive curvature admit a regular evolution domain
that avoids an unphysical singularity. In the matter dominated regime these
models are dynamicaly and observationaly indistinguishable from ``standard'' 
FLRW cosmology with a dust source.    
\end{abstract}


\section{Introduction.}
The  most general class of non-static, perfect fluid solutions of Einstein's
equations that are conformally flat is known as the``Stephani
Universe''\cite{KSMH},\cite{krasbook}. These mathematicaly simple solutions are
characterized (in general) by a shear-free but expanding and accelerating
4-velocity, with perfect fluid FLRW spacetimes being the acceleration free
subcase. The sphericaly symmetric Stephani
Universes\footnote{Unless specificaly stated otherwise, all further mention of
``Stephani Universes'' will refere to sphericaly symmetric Stephani Univeses
excluding their FLRW particular subclass} and some of their subcases have been examined in numerous papers (see
\cite{krasbook} for a comprehensive review). For example,  as star models
\cite{bofa},\cite{TW_1},\cite{TW_2},\cite{nato},\cite{bondi} and
as cosmologies generalizing FLRW\cite{krasbook},\cite{hen},\cite{pdl}. Other
papers have looked at their global properties and
singularities\cite{cook},\cite{kras83},\cite{suss88_1},\cite{suss88_2},
\cite{suss89},\cite{dabr93}, the thermodynamics of their fluid source
\cite{bonacol},\cite{qs},\cite{kqs}, and, more recently, as models with
inhomogeneous pressure in which to test recent astronomical data related to
supernova observations\cite{dabr95},\cite{dabr98}. Although these solutions  are
popular because of their mathematical simplicity, they are not considered 
physicaly meaningful because their fluid source seems quite artificial
and excessively restrictive: matter-energy density,
$\rho$, depends only on $t$ while pressure,
$p$, depends on $t$ and $r$ (in comoving coordinates). Another objection 
to the Stephani Universes is their incompatibility with a barotropic
relation\cite{mans},\cite{maspar} of the form
$p=p(\rho)$. However, the lack of fulfilment of barotropic
relations is not a strong argument to dismiss a given fluid
solution, since the latter might be compatible with more meaningful
non-barotropic equations of state.

The thermodynamics of the fluid source of the Stephani Universes is given by the
balance  equations of energy, momentum and entropy per particle, as well as by
the particle number conservation and the equilibrium Gibbs equation.
Previous papers have shown that the general Stephani Universe, having as source
a single component perfect fluid, is incompatible with the integrability
conditions of the Gibbs equation\cite{bonacol},\cite{kqs}, so that a ``
thermodynamical scheme'' is not possible (or in other words, a mathematicaly
consistent and closed set of thermodynamical equations of state is incompatible
with the field equations). However, this restriction does not apply to the
sphericaly symmetric case, or to any subcase whose isometry group has orbits
of dimension 2 or larger, hence mathematicaly simple equations of state have
been found\cite{qs} satisfying energy conditions, but without any basis on actual
physical matter models\footnote{The subclass of Stephani Universes studied in ref
\cite{qs} was erroneously reported as lacking isometries when it is sphericaly
symmetric. See \cite{krasbook} and \cite{kqs}}.  

Accepting the above mentioned limitations, this paper aims at looking for a
reasonable (or at least, less objectionable) interpretation for the Stephani
Universes. Using the Gibbs equation as a definition of temperature, $T$, it is
posible to determine the four free functions by imposing the constraint:
$p=nk_{_B}T$, where $n$ is particle number density and $k_{_B}$ is Boltzmann's
constant. However, the fulfilment of the relation between $\rho$ and $p$
characteristic of the equation of state of the ideal gas\cite{rkt} (together
with $p=nk_BT$) cannot hold because $\rho=\rho(t)$ while $p=p(t,r)$.  We
consider the equations of state of: (A) a single component classical monatomic
ideal gas, and (B) a mixture of non-relativistic and relativistic ideal gases,
where the internal energy of the former has been ignored (hence, $p$ and $T$ are
the radiation pressure and temperature). By writing up the explicit form of the
state variables for the forms of the free functions that satisfy $p=nk_BT$, we
provide explicitly the conditions for a tight approximation to the full
equations of state (A) and (B) in terms of the convergence of a series of powers
of adimensional products of
$\rho$ and a parameter $\varepsilon_0$, proportional to the energy ratio
$k_B\hat T_0/mc^2$, where $\hat T_0$ is a suitable constant 
temperature value and $m$ is the particles mass. This series converges in a
matter dominated low temperature regime characterized by $\varepsilon_0\ll 1$,
for models having spacelike slices (hypersurfaces of constant $t$, orthogonal to
the 4-velocity) with constant positive curvature (like ``closed''
FLRW cosmologies).  Under this approximation, the evolution of the fluid in the
matter dominated regime becomes very close to a dust FLRW cosmology, to such a
degree that observational parameters $H_0,\,\Omega_0,\,q_0$ become practicaly
indistinguishable from those of the FLRW spacetime. The same models for which the
above mentioned series converges are the only ones admiting an evolution range
that avoids unphysical features, such as a ''Finite Density'' singularity and an
``Asymptoticaly deSitter
boundary''(see \cite{suss88_1} to \cite{dabr93}), and so allowing for
a regular initial Cauchy hypersurface. Models whose spacelike slices orthogonal
to the 4-velocity have zero or negative curvature lack a regular Cauchy
hypersurface, while the series governing the approximation to the ideal gas
equations of state (A) and (B) only converge for comoving observers near the
symmetry center
$r=0$. 

Considering their qualities and defects, the subclass of Stephani Universes
derived and discussed in the present paper provide a reasonable and theoreticaly
consistent interpretation for this simple and well established class of exact
solutions.               

\section{The sphericaly symmetric Stephani Universes.}  

The metric in comoving coordinates for the sphericaly symmetric Stephani
Universes is usualy given in such a way as to emphasize its resemblance with
FLRW cosmologies\cite{krasbook}. However, we shall use instead the following
alternative comoving coordinates: 

$$ ds^2=-U^2c^2dt^2+ L^2\left[dr^2+f^2\left( {d\theta ^2+\sin ^2\theta \kern
1pt d\varphi ^2} 
\right)\right]\eqno(1a)$$
$$U={{{1+\left( K-{R\kern 1pt K_{,R}} 
\right)F^2} \over {1+K\kern 1pt F^2}}},\qquad L={{R}\over { {1+K\kern 1pt
F^2} }}\eqno(1b) $$

\noindent where $R=R(t)$, $K=K(t)$ and $K_{,R}=K_{,t}/R_{,t}=d K/dR $, and the
functions $F(r),f(r)$ are given by the three possible combinations
$$\matrix{{f=r,\quad F=r/2}\cr
{f=\sin r,\quad F=\sin (r/2)}\cr
{f=\sinh r,\quad F=\sinh (r/2)}\cr
}\eqno(2)$$

\noindent
The transformation of the radial coordinate $r$, relating (1)-(2) with the usual
radial coordinate for the Stephani Universes, $\bar r$, is given by
$r=\int{d\bar r}/(1+k\bar r^2/4)$, where $k=0,\pm 1$. The matter tensor
associated with (1) is a perfect fluid

$$ T^{a b}=\rho u^a u^b +p h^{a b}\eqno(3)$$ 
$$u^a=\frac{1}{U} \delta^a_t       \eqno(4)$$

\noindent where $ h^{a b}=c^{-2}u^au^b+g^{a
b}$ and $\rho,\,p$ are matter energy density and equilibrium pressure. The
time coordinate in (1) has been selected so that the expansion
scalar $\Theta=u^a_{;a}$ is given by 
$$\frac{\Theta}{3}=\frac{R_{,t}}{R}    \eqno(5)$$  

\noindent
The fluid source of the Stephani Universes is shear-free, hence the remaining nonzero kinematic
scalar is the 4-acceleration: $\dot u_a\equiv u_{a;b}u^b$, whose explicit form 
for (1) is
$$\dot u_a=\frac{U_{,r}}{U}\delta _a^r=-{{RK_{,R}\,f} \over {2\left[ {1+KF^2}
\right]\left[ {1+\left( {K-RK_{,R}}
\right)F^2} \right]}}\;\delta _a^r\eqno(6)$$

A perfect fluid source like (2) satisfies energy-momentum balance:
$T^{ab}\,_{;b}=0$, particle number conservation $(nu^a)_{;a}=0$ and entropy
balance: $(nsu^a)_{;a}=0$, where $n,s$ are particle number density and entropy
per particle. These balance and conservation laws take the following forms for
(1) and (2) 

$$u_aT^{ab}\,_{;b}=\dot \rho +(\rho+p)\Theta=0\eqno(7a)$$ 
$$h_{ab}T^{bc}\,_{;c}=h_a\,^bp_{,b}+(\rho+p)\dot u_a=0\eqno(7b)$$
$$(nu^a)_{;a}=\dot n+n\Theta=0\eqno(7c)$$
$$ (nsu^a)_{;a}=(\dot n+n\Theta)s+n\dot s=0\quad\Rightarrow\quad \dot s=0
    \eqno(7d)$$

\noindent
where $\dot A=u^a\,A_{,a}=u^t\,A_{,t}$ is the proper time derivative for any
scalar function $A$ and $\Theta,\,\dot u_a$ are given by (5) and (6). The state
variables must also comply with the equilibrium Gibbs equation

$$T\rm{d}s=\rm{d}\left(\frac{\rho}{n}\right)+p\rm{d}\left(\frac{1}{n}\right)\eqno(8)$$

\noindent
where $T$ is absolute temperature.

The field equation $G^t\,_t=(8\pi G/3c^2) T^t\,_t $ for (1), (2) and (3) is 

$${{8\pi G} \over {3c^2}}\;\rho =\left( {{{R_{,t}} \over R}}
\right)^2+{{\left( {K+k} \right)c^2} \over {R^2}}\eqno(9)$$

\noindent where:
$$k=\left\{ \matrix{\;\;0,\quad \quad f=r,\quad F=r/2,\hfill\cr
  \;\;1,\quad \quad f=\sin (r),\quad F=\sin (r/2),\hfill\cr
  -1,\quad \quad f=\sinh (r),\quad F=\sinh (r/2),\quad \hfill\cr}
\right.\eqno(10)$$

\noindent while the form of $p$ follows from (7a) as
$$p =-\rho -{R
\over 3}\frac{\rho _{,R}}{U}=-\rho -{R
\over 3}\rho _{,R}\left[ {{{1+K\kern 1pt F^2} \over
{1+\left( {K-R\kern 1pt K_{,R}}
\right)F^2}}} \right]    \eqno(11)$$

\noindent
The momentum ballance (7b) is

$$ p^\prime+{{fR^2\rho _{,R}\kern 1pt K_{,R}} \over 
{6\left[ {1+\left( {K-R\kern 1pt K_{,R}} \right)F^2}
\right]^2}}=0\eqno(12)$$

\noindent
where a prime denotes derivative wrt to $r$ and (6) and (11) were used to
eliminate $\dot u_a$ and $\rho+p$. The particle number density is given by
integrating (7c)

$$ n={{N(r)} \over {R^3}}\left[1+KF^2 \right]^3\eqno(13)$$

\noindent
where $N$ is the conserved particle number distribution.  Since the spacial
coordinates in (1) are comoving, the entropy balance (7d) yields simply $s=s(r)$,
and so, with the help of (7a) and (7c), the Gibbs equation (8) reduces to

$$Ts^\prime=(\rho+p)\left(\frac{1}{n}\right)^\prime\eqno(14)$$

\noindent
with $s'\ne 0$, since the fluid of Stephani Universes is not isentropic (or
equivalently, does not admit a barotropic relation $p=p(\rho)$).

\section{Determination of the free parameters.}
The  Stephani Universes contain (in general) two time dependent free functions
($R(t),\,K(t)$ in (1)) whose determination is elusive. The imposition of an
equation of state (ie a constraint relating state variables) should lead to the
determination of these free-parameters, howewer not all relations between the
state variables is compatible with these free functions and with the radial
dependence of the state variables (in particular, the barotropic relation
$p=p(\rho)$ is incompatible, see
\cite{mans},\cite{maspar}). Conversely, if the two free functions are determined
by suitable boundary conditions, or prescribed in order to obtain specific
kinematic effects, a specific constraint among state variables ensues, though
such constraint is a ``formal'' equation of state and might be wholy unphysical.
The usual boundary conditions invoked for determining the free functions are: a
matching with a Schwarzschild vacuum exterior along a comoving boundary and/or a
relation between $p$ and $\rho$ along a symmetry
center\cite{bofa},\cite{TW_1},\cite{TW_2},\cite{nato},\cite{bondi}, or the
desire to obtain specific kinematic efects, like bouncing \cite{bofa} or
acceleration of comoving observers\cite{dabr95},\cite{dabr98}, or a metric
element that is conformal to that of a FLRW
metric\cite{hen},\cite{pdl},\cite{suss89},\cite{dabr95},\cite{dabr98}. Besides
the time dependent free parameters, a proper thermodynamical approach produces
two extra free functions of $r$: the initial particle number distribution,
$N(r)$, and the entropy per particle $s(r)$.

The Stephani
Universes, as exact solutions subjected to the thermodynamical laws (7c), (7d)
and (8), become determined once the four free functions
$R(t),\,K(t),\,N(r),\,s(r)$ are either prescribed, found or obtained. From
equations (1)-(14), once $\rho$ and $K$ are known as functions of
$R$, the field equation (9) becomes the Friedmann equation whose integration
yields $R$ as a function of $t$. However, the functions $N(r),\,s(r)$ cannot be
determined from the geometry but from thermodynamical criteria.
The usual strategies to determine the free functions have lead, in all cases, to
equations of state whose physics has still remained unclear. We propose in
this paper another strategy for the determination of these free functions.

Consider the following relation associated with the equation of state of an ideal
gas: 
$$p=nk_{_B}T\eqno(15)$$
\noindent  where $k_{_B}$ is Boltzmann's constant. Using  $p,\,n$ from (11) and
(13), the fulfilment of (15) implies the constraint

$$k_{_B}T=\frac{k_{_B}(\rho+p)}{s'}\left(\frac{1}{n}\right)^\prime
=\frac{p}{n}\eqno(16a)$$

\noindent
which becomes explicitly
$$\left( {1+KF^2} \right)\left[ {\rho +{{R\rho _{,R}} \over 3}\left(
{1+{{k_BN'} \over {s'N}}} \right)} \right]+RF\left[ {{{2k_BF'} \over {s'}}K\rho
_{,R}-\rho K_{,R}F} \right]=0\eqno(16b)$$
\noindent
The only physicaly reasonable solution of (16) (other solutions yield
negative values of $n$ and $T$) follows if the terms in square parenthesis 
vanish independently. Since  
$\rho,\,K$ depend on $R$ while $F,\,N,\,s$ depend on $r$, we have
$${{R\rho _{,R}} \over {3\rho }}=-\left( {1+{{k_BN'} \over {s'N}}}
\right)^{-1}=a\eqno(17a)$$
$${{2k_BF'} \over {s'F}}={{K_{,R}/K} \over {\rho _{,R}/\rho }}=b\eqno(17b)$$
\noindent
where $a,\,b$ are arbitrary nonzero real constants. Equations (17) lead to      
$$\rho =\rho _0\left( {{{R_0} \over R}} \right)^{3a},\quad  \quad K=k_1\left(
{{\rho  \over {\rho _0}}} \right)^{b}=k_1\left( {{{R_0} \over R}}
\right)^{3ab}\eqno(18a)$$
$$\log N={{2(1-a)} \over b}\log F,\quad \quad s=s_0+{{ak_B}  \over {1-a}}\log
N,\quad \quad a\ne 1\eqno(18b)$$
$$N=N_0,\quad \quad \quad \quad \quad s=s_0+{{2k_B}  \over b}\log F,\quad \quad
\quad a=1\eqno(18c)$$

\noindent
Inserting (18) into (9), (11), (13) and using (15) leads to
$$\left( {{{R_{,t}} \over R}} \right)^2={{8\pi G} \over {c^2}}\rho _0\left(
{{{R_0} \over R}} \right)^{3a}-{{k_1c^2} \over {R^2}}\left( {{{R_0} \over R}}
\right)^{3ab}-{{k_0c^2} \over {R^2}}\eqno(19a)$$
$$p={{(a-1)\rho _0+\left[ {(1-3b)a-1} \right]\rho _0k_1\left( {R_0/R} 
\right)^{3ab}F^2} \over {1+(1+3ab)k_1\left( {R_0/R} \right)^{3ab}F^2}}\;\left(
{{{R_0} \over R}} \right)^{3a}\eqno(19b)$$
$$n={N \over {R^3}}\left[ {1+k_1\left( {{{R_0} \over R}} \right)^{3ab}F^2}
\right]^3\eqno(19c)$$
$$k_BT={{R_0^3} \over N}{{(a-1)\rho _0+\left[ {(1-3b)a-1} \right]\rho
_0k_1\left( {R_0/R} \right)^{3ab}F^2} \over {\left[ {1+k_1\left( {R_0/R}
\right)^{3ab}F^2} \right]^3\left[ {1+(1+3ab)k_1\left( {R_0/R} \right)^{3ab}F^2}
\right]}}\;\left( {{{R_0} \over R}} \right)^{3(a-1)}\eqno(19d)$$
From (18d), (18d), (19c) and (19d) it is evident that the choice $a\ne 1$ leads
to a diverging temperature along a symmetry center marked by $F(0)=0$. Hence, we
must consider only $a=1,\,b>0$, but in this case, in order to have positive
$p$ and $T$, we must also assume $k_1<0$. It turns to be convenient to redefine
the constant parameters $b$ and $k_1$ as
$$b=\gamma-1,\quad -k_1=\frac{\varepsilon_0}{3(\gamma-1)}$$ 
\noindent
where $\varepsilon_0$ is a positive constant. The metric functions in
(1), 4-velocity and 4-acceleration (4) and (6) become
$$U={{1-\varepsilon _0(\gamma _2/\gamma_1)F^2(R_0/R)^{\gamma_1}}  \over
{1-(\varepsilon _0/\gamma _1)F^2(R_0/R)^{\gamma_1}}},\quad L={R \over
{1-(\varepsilon _0/\gamma _1)F^2(R_0/R)^{\gamma_1}}}\eqno(20)$$
$$u^a={1 \over U}\delta ^a_t,\quad \dot u_a=\aleph\delta _a^r  ,\quad
\aleph={{{\textstyle{1 \over 2}}\varepsilon _0(R/R_0)^{\gamma_1}f} \over {\left[
{(R/R_0)^{\gamma_1}-(\varepsilon _0/\gamma _1)F^2} \right]\left[
{(R/R_0)^{\gamma_1}-\varepsilon _0(\gamma _2/\gamma_1)F^2} \right]}}\eqno(21)$$
where 
$$ \gamma _1=3(\gamma -1) ,\quad
\gamma _2=3\gamma -2$$
The state variables in (19) take the forms
$$\rho =\rho_0\left( {{{R_0} \over R}} \right)^3\eqno(22a)$$
$$ p={{\rho_0\varepsilon_0\;F^2} \over {1-\varepsilon
_0(\gamma_2/\gamma_1)\left( {\rho /\rho _0} \right)^{\gamma -1}F^2}}\left(
{{\rho 
\over {\rho _0}}} \right)^\gamma\eqno(22b)$$
$$ n={{N_0} \over {R_0^3}}\left( {{\rho  \over {\rho _0}}} \right)\left[
{1-\frac{\varepsilon _0}{\gamma_1}\left( {{\rho  \over {\rho _0}}}
\right)^{\gamma -1}F^2} \right]^3\eqno(22c)$$
$$k_BT={{\rho_0\varepsilon_0 R_0^3\;F^2} \over {N_0\left[
{1-(\varepsilon _0/\gamma_1)\left( {\rho /\rho _0} \right)^{\gamma -1}F^2}
\right]^3\left[ {1-\varepsilon _0(\gamma_2/\gamma_1)\left( {\rho /\rho _0}
\right)^{\gamma -1}F^2}
\right]}}\;\left( {{\rho  \over {\rho _0}}} \right)^{\gamma
-1}\eqno(22d)$$
while the Friedmann equation (9) is given by 
$$(R_{,t})^2={{8\pi G\rho_0R_0^3} \over
{3c^2R}}+\frac{\varepsilon_0c^2}{\gamma_1}\left( {{{R_0}
\over R}}
\right)^{\gamma_1}-kc^2\eqno(23)$$
This equation must be integrated in order to obtain $R=R(t)$ and have the models
fuly determined.

So far, we have determined the four free parameters of the Stephani Universe by
imposing the constraint (15). From (18c) and (22a), the state variables in
(22b-d) can be all be expressed as functions of two primary
thermodynamical variables $(\rho,s)$, so that rewriting (8) in terms of the
latter and using (15), it is straightforward to prove that the integrability
conditions of this Gibbs equation are fuly satisfied. However, the equations of
state of ideal  gases impose, besides (15), further relations among the
remaining state variables\cite{rkt}. We look at two of these equations of state
in the following section.

\section{Equations of state.}
The fulfilment of the relation $ p=nk_BT$ leads to the consideration of various
possible ideal gas equations of state for the fluid. However, we have:
$\rho=\rho(t)$, while the remaining state variables depend on both $t$ and
$r$, hence these equations of state will not hold exactly, though the constant
parameters $\gamma,\,\varepsilon_0,\,\rho_0$ can be determined in such a way that
the full equation of state holds under suitable approximations.  Two cases will
be examined separately below.

\subsection{The classical monatomic ideal gas}
For this case the equation of state is (15) together with\cite{rkt},\cite{susstr}
$$\rho=mc^2n+\frac{3}{2}n\,k_BT \eqno(24)$$
As mentioned before, an equation like (24) cannot be satisfied exactly. In order
to examine under which conditions it can be satisfied approximately, we use
(15), (22b) and (22c) to write the right hand side of (24) expanded up to third
order in
$\varepsilon_0$ 
$$mc^2n+{3 \over 2}nk_BT={{mc^2N_0} \over {R_0^3}}\left( {{\rho   \over {\rho
_0}}} \right)+\left[ {{3 \over 2}\rho _0-{{mc^2N_0} \over {(\gamma -1)R_0^3}}}
\right]\left( {{\rho  \over {\rho _0}}} \right)^\gamma F^2\varepsilon _0+\left[
{{{3\gamma -2} \over {2(\gamma -1)}}\rho _0+{{mc^2N_0} \over {3(\gamma
-1)^2R_0^3}}} \right]\left( {{\rho  \over {\rho _0}}} \right)^{2\gamma
-1}F^4\varepsilon _0^2
$$
$$+\left[ {{{(3\gamma -2)^2} \over {6(\gamma
-1)^2}}\rho _0-{{mc^2N_0} \over {27(\gamma -1)^3R_0^3}}} \right]\left( {{\rho 
\over {\rho _0}}} \right)^{3\gamma -2}F^6\varepsilon _0^3+O(\varepsilon
_0^4)\qquad\qquad\qquad\qquad\eqno(25)
$$
Since $\rho_0,\,R_0$ are $\rho,\,R$ evaluated at present cosmic time $t=t_0$
when matter dominated conditions are assumed to prevail, then energy density is
overwhelmingly dominated by rest mass energy, and so, looking for the best
possible approximation of (25) to (24), leads to the following identifications
in (25)   
$$\rho_0=\frac{mc^2N_0}{R_0\,^3},\qquad
\frac{1}{\gamma-1}=\frac{3}{2}\quad \Rightarrow\quad \gamma=\frac{5}{3}
\eqno(26)$$
Inserting (26) into (25) yields 
$$mc^2n+{3 \over 2}nk_BT=\rho \left[ {1+3\left( {{\rho  \over {\rho _0}}} 
\right)^{4/3}F^4\varepsilon _0^2+{{13} \over 4}\left( {{\rho  \over {\rho _0}}}
\right)^2F^6\varepsilon _0^3+O(\varepsilon _0^4)} \right]\eqno(27)$$
so that the leading nonzero order term in the series inside the square
brackets is quadratic in $\varepsilon_0$.  Comparing (27) with (24) shows that
(22) and (26) lead to a good approximation to (24) provided the power series
inside the square brackets in (27) converges. A sufficient condition for this
convergence requires that each term (with the exception of the zero order term)
must satisfy
$$\left( {{\rho  \over {\rho _0}}} \right)^{2\xi/3}F^{2\xi}(\varepsilon
_0)^\xi\ll1,\quad
\quad \xi=2,3,..\eqno(28)$$
The form of the adimensional parameter $\varepsilon_0$ follows by
using (22b), (22d) and (26) in order to expand $p$ and $k_BT$ around
$\varepsilon_0$ 

$$p=\frac{mc^2N_0\,\varepsilon_0\,F^2}{R_0^3}\left(\frac{\rho}
{\rho_0}\right)^{5/3}\left[1+\frac{3}{2}\left(\frac{\rho}
{\rho_0}\right)^{2/3}F^2\varepsilon_0+\frac{9}{4}\left(\frac{\rho}
{\rho_0}\right)^{4/3}F^2\varepsilon_0^2+O(\varepsilon_0^3)\right]
\eqno(29a)
$$
$$k_BT=mc^2\,\varepsilon_0\,F^2\left(\frac{\rho}
{\rho_0}\right)^{2/3}\left[1+3\left(\frac{\rho}
{\rho_0}\right)^{2/3}F^2\varepsilon_0+6\left(\frac{\rho}
{\rho_0}\right)^{4/3}F^2\varepsilon_0^2+O(\varepsilon_0^3)\right]
\eqno(29b)
$$
By looking at the common factor outside the square brackets, it becomes natural
to identify
$$ mc^2\,\varepsilon_0=k_B\hat T_0\qquad \Rightarrow\qquad
\varepsilon_0=\frac{k_B\hat T_0}{mc^2}
\eqno(30)$$
where $\hat T_0$ is a suitable constant temperature value. With the help of
(22), (26) and (30), the exact form of the state variables becomes
$$p={{(N_0k_B\hat T_0/R_0^3)\;F^2} \over {1-{\textstyle{{3} \over
{2}}}\varepsilon _0\left( {\rho /\rho _0} \right)^{2/3}F^2}}\left( {{\rho 
\over {\rho _0}}} \right)^{5/3} ={{(N_0k_B\hat T_0/R_0^3)\;F^2} \over
{1-{\textstyle{3 \over {2}}}\varepsilon _0\left( {R_0/R}
\right)^2 F^2}}\left( {{{R_0} \over R}} \right)^5\eqno(31a)$$
$$n={{N_0} \over {R_0^3}}\left( {{\rho  \over {\rho _0}}} \right)\left[
{1-{{\varepsilon _0} \over {2}}\left( {{\rho  \over {\rho _0}}}
\right)^{2/3}F^2} \right]^3={{N_0} \over {R^3}}\left[
{1-{{\varepsilon _0}
\over {2}}\left( {{{R_0} \over R}} \right)^2F^2}
\right]^3\eqno(31b)$$
$$k_BT={{k_B\hat T_0\;F^2} \over {\left[ {1-{\textstyle{{\varepsilon _0}
\over {2}}}\left( {\rho /\rho _0} \right)^{2/3}F^2}
\right]^3\left[ {1-{\textstyle{{3} \over {2}}}\varepsilon _0\left( {\rho /\rho
_0} \right)^{2/3}F^2}
\right]}}\;\left( {{\rho  \over {\rho _0}}} \right)^{2/3}    
       ={{k_B\hat T_0\;F^2} \over {\left[ {1-{\textstyle{{\varepsilon _0} 
\over {2}}}\left( {R_0/R} \right)^2F^2} \right]^3\left[
{1-{\textstyle{{3} \over {2}}}\varepsilon _0\left( {R_0/R} \right)^2F^2}
\right]}}\;\left( {{{R_0} \over R}} \right)^2\eqno(31c)$$
while the Friedmann equation (22d) is given by
$$(R_{,t})^2={{8\pi GmN_0} \over {3R}}+{{\varepsilon _0c^2} \over {2}}\left(
{{{R_0} \over R}} \right)^2-kc^2\eqno(32)$$
 
The degree in which these state variables provide a good approximation to the
equation of state (15) (24) strongly depends on the behavior of $F$,
$\rho/\rho_0$ and on the value of $\varepsilon_0$ in (30), a small number for an
enormous range of temperatures (if $m$ is a protonic mass, we have
$\varepsilon_0\approx 10^{-4}$ for as high as $\hat T_0\approx 10^8$ degrees K.).
The fulfilment of (28) for specific energy and temperature ranges of the models
will be discussed in section VII.

\subsection{Mixture of matter and radiation}
Consider a mixture of two ideal gases, one non-relativistic (superindex ``I'')
and the other ultra-relativistic (superindex ``II''), the equation of state in
this case is
$$\rho=mc^2n^{\rm{I}}+\frac{3}{2}n^{\rm{I}}k_BT^{\rm{I}}
+3n^{\rm{II}}k_BT^{\rm{II}}\eqno(33a)$$ 
$$p=n^{\rm{I}}k_BT^{\rm{I}}+n^{\rm{II}}k_BT^{\rm{II}}\eqno(33b)$$
If the internal energy of the non-relativistic gas can be neglected:
$n^{\rm{I}}k_BT^{\rm{I}}\ll n^{\rm{II}}k_BT^{\rm{II}}$ but (at least at a
given stage of the evolution) the energy density $n^{\rm{II}}k_BT^{\rm{II}}$ is
not negligible in comparison with the rest mass energy $mc^2n^{\rm{I}}$, then
(33) can be approximated by\cite{susstr},\cite{susspav}
$$\rho= mc^2n^{\rm{I}}+3n^{\rm{II}}k_BT\eqno(34a)$$ 
$$p=n^{\rm{II}}k_BT\eqno(34b)$$
\noindent
where $T=T^{\rm{II}}$ and the particle number densities independently satisfy
conservation laws like (7c), therefore $n^{\rm{I}},\,n^{\rm{II}}$ are given by
expressions identical to (22c) with $N_0$ replaced by
$N_0^{\rm{I}}$ and $N_0^{\rm{II}}$. The approximation (34a-b) can be especialy
suited for a mixture of photons and barions\cite{susspav} bacause the ratio of
the former to the latter is such a large number ($\approx 10^8$), leading to a
reasonable description of cosmological mixtures of non-relativistic matter and
radiation for energy and temperature ranges prevailing after cosmological
nucleosynthesis, including the radiative era (recombination and decoupling), and
up to the present. 

As in the previous subsection, (34a) cannot hold exactly but can be aproximated
expanding its right hand side in terms of $\varepsilon_0$. From the forms of
$n^{\rm{I}}$ and $T=T^{\rm{II}}$\,from (22c) and (22d),
and using (34b) in the form $3n^{\rm{II}}k_BT=3p$ with $p$ given by (22b),\, we
obtain  
$$mc^2n^{\rm{I}}+3n^{\rm{II}}k_BT={{mc^2N_0^{\rm{I}}} \over {R_0^3}}\left(
{{\rho   \over {\rho _0}}} \right)+\left[ {3\rho _0-{{mc^2N_0^{\rm{I}}} \over
{(\gamma -1)R_0^3}}}
\right]\left( {{\rho  \over {\rho _0}}} \right)^\gamma F^2\varepsilon _0+\left[
{{{3\gamma -2} \over {(\gamma -1)}}\rho _0+{{mc^2N_0^{\rm{I}}} \over {3(\gamma
-1)^2R_0^3}}} \right]\left( {{\rho  \over {\rho _0}}} \right)^{2\gamma
-1}F^4\varepsilon _0^2
$$
$$+\left[ {{{(3\gamma -2)^2} \over {3(\gamma
-1)^2}}\rho _0-{{mc^2N_0^{\rm{I}}} \over {27(\gamma -1)^3R_0^3}}} \right]\left(
{{\rho 
\over {\rho _0}}} \right)^{3\gamma -2}F^6\varepsilon _0^3+O(\varepsilon
_0^4)\qquad\qquad\qquad\qquad\eqno(35)
$$
so that, considering matter dominated conditions and aiming at the best possible
approximation, we identify 
$$\rho_0=\frac{mc^2N_0^{\rm{I}}}{R_0\,^3},\qquad
\frac{1}{\gamma-1}=3\quad \Rightarrow\quad \gamma=\frac{4}{3}
\eqno(36)$$
transforming (35) into

$$mc^2n^{\rm{I}}+{3 \over 2}n^{\rm{II}}k_BT=\rho \left[ {1+9\left( {{\rho  \over
{\rho _0}}} 
\right)^{2/3}F^4\varepsilon _0^2+11\left( {{\rho  \over {\rho _0}}}
\right)^{4/3}F^6\varepsilon _0^3+O(\varepsilon _0^4)} \right]\eqno(37)$$

\noindent
indicating, as in the previous subsection, that a good approximation to (34a)
requires the convergence of the power series inside the square brackets, leading
to a condition very similar to (28)
$$\left( {{\rho  \over {\rho _0}}} \right)^{\xi/3}F^{2\xi}(\varepsilon
_0)^\xi\ll1,\quad
\quad \xi=2,3,..\eqno(38)$$
The form of $\varepsilon_0$ follows, as in the previous subsection,  by obtaining
with the help of (22b), (22d) and (36) the forms of $p$ and $k_BT$ expanded in
terms of $\varepsilon_0$, leading  to
$$p=\frac{mc^2N_0^{\rm{I}}\,\varepsilon_0\,F^2}{R_0^3}\left(\frac{\rho}
{\rho_0}\right)^{4/3}\left[1+2\left(\frac{\rho}
{\rho_0}\right)^{1/3}F^2\varepsilon_0+4\left(\frac{\rho}
{\rho_0}\right)^{2/3}F^2\varepsilon_0^2+O(\varepsilon_0^3)\right]
\eqno(39a)
$$
$$k_BT=\frac{mc^2N_0^{\rm{I}}\,\varepsilon_0\,F^2}{N_0^{\rm{II}}}
\left(\frac{\rho} {\rho_0}\right)^{1/3}\left[1+5\left(\frac{\rho}
{\rho_0}\right)^{1/3}F^2\varepsilon_0+16\left(\frac{\rho}
{\rho_0}\right)^{2/3}F^2\varepsilon_0^2+O(\varepsilon_0^3)\right]
\eqno(39b)
$$
so that it becomes natural to identify
$$ mc^2N_0^{\rm{I}}\,\varepsilon_0=N_0^{\rm{II}}k_B\hat
T_0,\qquad\Rightarrow\qquad \varepsilon_0=\frac{N_0^{\rm{II}}k_B\hat
T_0}{N_0^{\rm{I}}mc^2}    
\eqno(40)$$
where $\hat T_0$ is a suitable constant temperature value for the relativistic
component. Using (22), (23), (36) and (40), the state variables take the
following exact forms
$$p={{(N_0^{\rm{II}}k_B\hat T_0/R_0^3)\;F^2} \over {1-2\varepsilon
_0\left( {\rho /\rho _0} \right)^{1/3}F^2}}\left( {{\rho 
\over {\rho _0}}} \right)^{4/3} ={{(N_0^{\rm{II}}k_B\hat T_0/R_0^3)\;F^2} \over
{1-2\varepsilon _0\left( {R_0/R}
\right) F^2}}\left( {{{R_0} \over R}} \right)^4\eqno(41a)$$
$$n={{(N_0^{\rm{I}}+N_0^{\rm{II}})} \over {R_0^3}}\left( {{\rho  \over
{\rho _0}}}\right)\left[ {1-\varepsilon _0F^2\left( {{\rho  \over {\rho _0}}}
\right)^{1/3}} \right]^3={{(N_0^{\rm{I}}+N_0^{\rm{II}})} \over {R^3}}\left[
{1-\varepsilon _0F^2\left( {{{R_0} \over R}} \right)}
\right]^3\eqno(41b)$$
$$k_BT={{k_B\hat T_0\;F^2} \over {\left[ {1-\varepsilon _0F^2\left( {\rho
/\rho _0} \right)^{1/3}}
\right]^3\left[ {1-2\varepsilon _0F^2\left( {\rho /\rho
_0} \right)^{1/3}}
\right]}}\;\left( {{\rho  \over {\rho _0}}} \right)^{1/3}    
       ={{k_B\hat T_0\;F^2} \over {\left[ {1-\varepsilon _0F^2\left(
{R_0/R}\right)} \right]^3\left[ {1-2\varepsilon _0F^2\left(
{R_0/R} \right)}
\right]}}\;\left( {{{R_0} \over R}} \right)\eqno(41c)$$
$$(R_{,t})^2={{8\pi GmN_0^{\rm{I}}} \over {3R}}+\varepsilon _0c^2\left( {{{R_0}
\over R}} \right)-kc^2\eqno(42)$$

The fulfilment of (38) with $\varepsilon_0$ given by (40) controls how close
the state variables aproximate (34a). However in the present case the ratio
$N_0^{\rm{II}}/N_0^{\rm{I}}$ must be taken into consideration as well. This
will be discussed in section VII. Independently of the equations of state
presented so far, the evolution range of the models is necessarily restricted by
singularities and other geometric features. We examine these features in the
following section.

\section{Singularities, regularity domain, symmetry centers and FLRW limit}

Looking at $\rho$ $p$ and $T$ in equations (22), (31) and (41), pending the
determination of $R=R(t)$ by integrating (32) and (42), we can identify the
following scalar curvature singularities
$$R=0\eqno(43a)$$
$$1-\frac{3\gamma-2}{3(\gamma-1)}\varepsilon_0F^2
\left(\frac{R_0}{R}\right)^{3(\gamma-1)}=0\eqno(43b)$$
\noindent
The singularity (43a) is marked by a hypersurface $t=\hbox{const}$, all state
variables diverge and is similar to a ``big-bang'' in FLRW spacetimes.
The type (43b) has been identified in previous literature as a ``finite
density'' (FD) singularity\cite{suss88_1},\cite{suss88_2}. It is, in general, a
hypersurface marked by a curve in the
$(t,r)$ or $(R,r)$ plane, though it is an unphysical singularity because $p,\,T$
diverge while $\rho$ remains finite
$$\left[\frac{\rho}{\rho_0}\right]_{\rm{FD}}=\left[\frac{3(\gamma-1)}{(3\gamma-2)
\varepsilon_0F^2}\right]^{1/(\gamma-1)}\eqno(44)
$$ 
for all comoving observers, except along the center worldline $r=0$ where
$F(0)=0$ and so $[\rho/\rho_0]_{\rm{FD}}$ diverges. Particle number density also
remains finite, as can be verified by inserting (44) into (22c).

Another feature worth remarking is the points in spacetime where
$$1-\frac{\varepsilon_0}{3(\gamma-1)}F^2\left(\frac{R_0}{R}
\right)^{3(\gamma-1)}=0\eqno(45)$$ 
These points mark a spacetime boundary
characterized by an asymptoticaly inflationary or (asymptoticaly de
Sitter, ``ADS'') behavior\cite{suss88_2}: $p+\rho\to 0$ and
$n\to 0$. Although $k_BT=p/n$ diverges along (45) (because $n$ vanishes for
finite $p$), this is not realy a curvature singularity since all polynomial
curvature scalars are bounded. 

Since for the cases we are interested ($\gamma=4/3,\,5/3$), the FD singularity
is marked by larger values of $R/R_0$ than the ADS boundary, and since this
singularity is spacelike \cite{suss88_1}\cite{suss88_2}, we will consider as the
regular evolution domain of the solutions all points of spacetime along the
fluid worldlines lying to the future of the FD singularity. These domain can be
characterized, for every comoving observer, by either one of the following
conditions

$$\frac{R}{R_0}>\left[\frac{(3\gamma-2)\varepsilon_0F^2}{3(\gamma-1)
}\right]^{1/(3\gamma-3)},\qquad
\frac{\rho}{\rho_0}<\left[\frac{3(\gamma-1)}{(3\gamma-2)
\varepsilon_0F^2}\right]^{1/(\gamma-1)}\eqno(46)
$$

From the material presented in previous sections, it is obvious that the
FLRW limit of the Stephani Universes is given by $\varepsilon_0=0\Rightarrow K=
0$ and
$N=N_0$, so that
$\dot u_a=p'=s'=0$. Since $\varepsilon_0=0$ implies $p=T=0$, this
limit leads to a dust FLRW universe, and so a sort of  
asymptotic dust FLRW state can be reached as the fluid evolves in
time. Also, for whatever form of $R(t)$ obtained from integrating (32) and
(42), the worldlines marked by $r=r_c$ where $f(r_c)=0$, with $f$
given by (2) and (10), characterize the symmetry centers. Along these worldlines,
the metric function $\sqrt{g_{\theta\theta}}=Lf$, the radius of the orbits of the
rotation group, vanishes regularly. Also, $\dot u_a,\,p'$ vanish, as well the
radial gradients of all other geometric or thermodynamical variables should also
vanish, hence spacetime appears isotropic for observers along symmetry centers.
Depending on the functions in (2) and (10), the Stephani Universes admit one
($r=0$ for $k_0=0,-1$) or two ($r=0,\pi$ for $k_0=1$) symmetry centers.

The singularities (43a-b) and regular boundary (45) have been
examined extensively elsewhere, and so we refer the reader to the appropriate
references (see \cite{suss88_1} to \cite{dabr93}). Considering (46) for 
the cases $k=0,-1$ of $F$ in (2) and (10), it is evident that
these cases do not allow for the existence of a hypersurface
$t=\hbox{const}$ in the regularity domain that is also a regular Cauchy surface.
In the case $k=1$ we have two symmetry centers marked by $r=0$ and
$r=\pi$, and so the full domain of the radial coordinate is $0\le r\le \pi$.
Since in this case we have $0\le F=\sin(r/2)\le 1$, it follows from (46) that a
regular Cauchy hypersurface $t=t_i$ can be found in the regularity domain, so
that a regular evolution takes place in the range $t_i<t<t_M$, where
$t_M$ marks the hypersurface of maximal expansion characterized by $\Theta=0$.
In this case, the regular hypersurfaces $t=\hbox{const}>t_i$ will be 
diffeomorphic to the invariant spacelike slices of a ``closed'' FRLW spacetime.

\section{Observational features}
As shown in section IV, the fluid sources of the models approximately
satisfy the ideal gas equations of state (15),(24) and (34), provided conditions
such as (26), (28), (36) and (38) hold. The evolution of the models towards the
present time ($t\to t_0$, $R\to R_0$) approaches that of a FLRW dust cosmology,
characterizing a homogeneous and isotropic matter dominated universe. It is not
surprising that (15), (24) and (34) approach dust sources, the latter being realy
a convenient approximation of ideal gases in a low temperature and pressure
limits. However, the ``nearly dust FLRW'' approximation of the models must be
examined by verifying how the observational parameters approach as $t\to
t_0$ those associated with FLRW cosmologies. 

The resemblance of the field equation (9) with the Friedmann equation of a FLRW
spacetime, together with the fact that $\rho=\rho(t),\,\Theta=\Theta(t)$, leads
to an invariant definition of observational parameters $H=\Theta/3=\dot
L/L=R_{,t}/R$ and $\Omega=8\pi G/(3c^2H^2)$ that is also very similar to their
FLRW equivalents. Evaluating along $R=R_0$, we have

$$H_0^2=\frac{8\pi
G}{3c^2}\rho_0+\left[\frac{\varepsilon_0}{\gamma_1}
-k\right]\frac{c^2}{R_0^2}\eqno(47a)
$$
$$\Omega_0\equiv \frac{8\pi G \rho_0}{3c^2H_0^2}=\frac{8\pi G\rho_0
R_0^2/3c^4}{8\pi G\rho_0 R_0^2/3c^4+(\varepsilon_0/\gamma_1)-k}\eqno(47b)
$$
$$ 
\Omega_0-1=\left(k-\frac{\varepsilon_0}{\gamma_1}\right)\frac{c^2}{(H_0R_0)^2}
\eqno(47c) 
$$
generalizing the FLRW parameters
$$\tilde H_0^2 =\frac{8\pi
G}{3c^2}\rho_0-k\frac{c^2}{R_0^2},\qquad \tilde\Omega_0=\frac{8\pi G
\rho_0}{c^2\tilde H_0^2}=\frac{8\pi G\rho_0 R_0^2/3c^4}{8\pi G\rho_0
R_0^2/3c^4-k}\eqno(48)$$ 
where $\gamma_1=3(\gamma-1)$ and all quantities with tildes will henceforth be
FLRW quantities. Since we are interested in comparing observations in the
Stephani Universes and in FLRW dust cosmologies at $R=R_0$, it is useful to
expand $H_0$ and $\Omega_0$ in terms of $\tilde H_0,\,\tilde\Omega_0 $ up to
first order in $\varepsilon_0$

$$H_0\approx\tilde H_0\left[1+\frac{c^2}{\gamma_1(\tilde H_0
R_0)^2}\varepsilon_0+O(\varepsilon_0^2)\right],\qquad \Omega_0\approx 
\tilde\Omega_0\left[1-\frac{c^2}{\gamma_1(\tilde H_0
R_0)^2}\varepsilon_0+O(\varepsilon_0^2)\right],\qquad
\tilde\Omega_0-1=\frac{kc^2}{(\tilde H_0 R_0)^2}\eqno(49)
$$ 
showing that, in general $H_0>\tilde H_0 $ and $\Omega_0<\tilde\Omega_0$, and
that
$H_0,\,\Omega_0$ are roughly equal to $\tilde H_0,\, \tilde\Omega_0$ plus
corrections of the order
$\approx\varepsilon_0\tilde\Omega_0\approx\varepsilon_0$. Another important
observational quantity is the deceleration parameter
$\tilde q$, defined for FLRW cosmologies as
$$\tilde q_0\equiv -\left[\frac{R_{,tt}R}{R_{,t}^2}\right]_0=
\frac{1}{2}\tilde\Omega_0$$
We can compute, with the help of $H$ given by (9) for the Stephani Universes, an
equivalent quantity $q_0$ and expand it in terms of $\tilde q_0$
$$q_0=-\left[\frac{H_{,t}+H^2}{H^2}\right]_0=\frac{1}{2}\frac{8\pi G\rho_0
R_0^2/3c^4+\varepsilon_0}{8\pi
G\rho_0R_0^2/3c^4+\varepsilon_0/\gamma_1-k}=\frac{\Omega_0}{2}
\frac{2+\varepsilon_0}{2-\varepsilon_0}-\frac{\varepsilon_0}{2-\varepsilon_0}
\eqno(50a)$$
$$q_0\approx\tilde
q_0+\frac{1}{2}\left[1-\frac{\tilde\Omega_0}{\gamma_1}\right]\frac{c^2}
{(\tilde H_0R_0)^2}\,\varepsilon_0+O(\varepsilon_0^2)\eqno(50b) $$
However the definition of $q$, unlike $H,\,\Omega$, is coordinate dependent. In
order to derive the appropriate equivalent of $\tilde q_0$ for the models based
on the Stephani Universes, we need to derive the general form of the
relation between red shift and luminosity distance for local
observations along
$t=t_0,\,r=r_0$.  

Consider the simplest form of a radial null
vector
$k^a=(k^t,k^r)$  for the metric (1) with $U$ and $L$ given by (20)

$$k^t=\frac{cdt}{dv}=\frac{L}{U}=\frac{R}{1-\varepsilon _0\gamma
_2F^2(R_0/R)^{3(\gamma -1)}},\qquad k^r=\frac{dr}{dv}=\pm
1,\qquad u_ak^a=-L\eqno(51)$$ 
The effects of inhomogeneity and anisotropy on red shift observations in $t=t_0$
of nearby sources ($z\ll 1$) can be appreciated from the relation between
luminosity distance vs red shift given for the radial light rays (51) by
\cite{elismac}
$$z=\left[ {u_{a;b}\hat k^a\hat k^b} \right]_0\Delta \ell  _0+{1 \over 2}\left[
{u_{a;bc}\hat k^a\hat k^b\hat k^c} \right]_0\Delta \ell _0^2+O(\Delta \ell
_0^3)\eqno(52a)$$
$$\hat k^a=\left( {u_bk^b} \right)^{-1}k^a,\quad \hat k^t=-U^{-1}=-u^t,\quad 
\hat k^r=\mp L,\qquad \Delta \ell _0=-\left[ {u_bk^b} \right]_0\Delta
\lambda=L_0 \Delta \lambda\eqno(52b)$$
where $\Delta \lambda$ is the geodesic parameter of an observed gallaxy. For an
expanding, accelerating but shear-free 4-velocity, the quantities involved in
(52) are
$$u_{a;b}\hat k^a\hat k^b={\Theta  \over 3}-\dot u_a\hat k^a\eqno(53a)$$
$$u_{a;bc}\hat k^a\hat k^b\hat k^c={{\Theta _{,c}} \over 3}\hat k^c+{{2\Theta
^2} \over 9}-\Theta \dot u_a\hat k^a-\dot u_{b;c}\hat k^b\hat k^c+\dot u_a\dot
u_b\hat k^b\hat k^c\eqno(53b)$$
Evaluating these expressions for (20), (21), (48b) and (47), expanding up to
first order in $\varepsilon_0$ and evaluating at $(t_0,r_0)$ yields
$$z\approx{{H_0} \over c}\left[ {1\mp {{f_0c} \over {2H_0R_0}}\varepsilon _0}
\right]\Delta \ell _0+{{H_0^2} \over {2c^2}}\left\{ {3+q_0+\varepsilon _0\left[ {\mp (1+q_0)F_0^2+
{{(\gamma _1+5)f_0c} \over {2H_0R_0}}\mp {{(1-2kF_0^2)c^2} \over {2(H_0R_0)^2}}}
\right]} \right\}\Delta \ell _0^2+O(\Delta \ell _0^3)\eqno(54)
$$
where $H_0,\,q_0$ are given by (47a) and (50). As shown by (54), the red shift
observed from a given source depends in general on the position of the observer
(its radial coordinate $r_0$ determining $f_0,F_0$). This is expected, since the
metric given by (1) and (20) has an explicit radial dependence contained in the
functions $f=4FF'$, and so the red shift distribution is anisotropic because of
the inhomogeneity of the cosmological model. In particular, an
accelerating and shear-free fluid leads to a monopole plus dipole term in (53a),
since the direction of the 4-acceleration ($\partial/\partial r$) is a
privileged direction. Hence, for a connecting vector in the radial
direction $\Delta\ell_0$ has a maximal value, while along other directions it
must be multiplied by a factor $\cos\Psi$, where $\Psi$ is the telescopic
angle\cite{elismac}. Observations along the symmetry center
$r=0$ detect an isotropic distribution
  
$$z|_{r=0}=\frac{H_0}{c}\Delta\ell_0+\frac{H_0^2}{2c^2}\left[3+q_0\mp
\varepsilon_0\frac{c^2}{2(H_0R_0)^2}\right]\Delta\ell_0^2+O(\Delta\ell_0^3)
\eqno(55)$$
Equation (54) provides then the maximal
magnitude of the corrections to local red shift isotropy that inhomogeneity of
spacetime introduces into to the red shift vs luminosity distance.  From (47c)
and since $f_0\approx F_0$, these corrections are of the
order $\varepsilon_0\Omega_0F_0\approx
\varepsilon_0\bar\Omega_0F_0\approx \varepsilon_0F_0$. For these corrections to
be small, we need $\varepsilon_0F_0\ll 1$, which for the models considered in
sections III and IV, means that either $k=1$ (where $0\le r\le \pi$ and $f,F$
are finite), or if $k=0,\,-1$, the domain $r$ of $f,F$ must be necessarily
restricted. However, as discussed in section V, the cases
$k=0,-1$ do not allow for regular hypersurfaces $t=\hbox{const}$ nor a regular
initial Cauchy surface at $t=t_0$. For the case
$k=1$, if $\varepsilon_0\ll 1$ the corrections due to inhomogeneity are small
for all $r_0$ compared with the isotropic zero order terms $H_0,\,q_0$, which
means that local observations are overwhelmingly determined by $H_0,\,q_0$,
leading to a distribution of red shifts that is almost isotropic but whose
observational parameters slightly differ from the FLRW parameters $\tilde
H_0,\,\tilde q_0$. However, because of (49) and (50), $H_0,\,q_0$ differ from
$\tilde H_0,\,\tilde q_0$ by corrections of order $\varepsilon_0$, and so for
sufficiently small
$\varepsilon_0$, compatible with (28) and (38), this difference should be much
smaller than the 10-15 percent or so observational uncertainty\cite{peacock} in
the most accurate possible local measurements of $\tilde H_0$. This will be
discussed in the following section.    

\section{Discussion and conclussion.}
We have presented in previous sections important general features of the 
subclass of Stephani Universes derived in sections III and IV. We still need
to prove that the approximation to ideal gas equations of state (conditions (28)
and (38)), as well as the approximation to isotropic local observations, are
actualy met for physicaly reasonable numerical values of the parameters
involved. We also need to integrate the Friedmann equations (32) and (42) in
order to obtain a fuly determined dynamical evolution of the models, allowing us
to verify their prediction of the ``age of the universe today'' in comparison
with the prediction of standard cosmology (dust FLRW universes). Each case,
$\gamma=4/3$ and $\gamma=5/3$, is examined separately below.

\subsection{The classical monatomic ideal gas source.}

For this model the state variables $\rho,p,n,T$ are given by (22a) and (31a-c),
satisfying (15) and providing (if (28) holds) a close approximation to (24). 
Consider matter dominated conditions along $t=t_0$, so that $\rho_0$ is
matter-energy density today. In order to examine the fulfilment of (28), 
suppose that $m$ is a proton mass and (say) $\hat T_0\approx 100$ degrees K,
leading to $\varepsilon_0\approx 10^{-10}$ in (30). The terms of the form (28) in
the series in (27) have the form    

$$\left(\frac{\rho}{\rho_0}\right)^{2\xi/3}F^\xi\,\varepsilon_0^\xi=F^\xi\,
\left(\frac{\rho} {\rho_0}\right)^{2\xi/3}
\left(\frac{k_B\hat T_0}{mc^2}\right)^\xi\approx
\left(10^{-10}\right)^\xi F^\xi\,\left(\frac{\rho} {\rho_0}\right)^{2\xi/3}\ll
1,\qquad \xi=2,3,..\eqno(56)$$

\noindent controling the goodness of the approximation of (27) to (24) along the
directions of increasing $r$ (increasing $F$) and $t$ (decreasing $\rho$).  
For the cases $k=0,-1$, the function $F$ increses monotonously as $r$ increases,
so that even for very small values of $\varepsilon_0$ and $\rho/\rho_0$ we might
have a diverging series in (27), since $F^\xi<F^{\xi+1}$ for $r>2$. Hence in the
cases $k=0,-1$, a necessary (but not sufficient) condition for (27) to converge
into (24) is $r<2$, restricting the domain of regularity to comoving observers
near the center worldline $r=0$. In the case $k=1$, the function $F$ is bounded
and satisfies $F^\xi>F^{\xi+1}$ for all values of $\xi$ and $0\le r\le \pi$.
Therefore, in this case the convergence of the series in (27) and the fulfilment
of (56) only depends on the comparison of $\varepsilon_0$ with ratio
$\rho/\rho_0$ for the leading term of nonzero order ($\xi=2$) in (27). From (46)
and using the value
$\varepsilon_0\approx 10^{-10}$, the allowed values of $\rho/\rho_0$ (with the
maximal value at the FD density) is given by

$$\frac{\rho}{\rho_0}<\left[\frac{2}{3\varepsilon_0F^2}\right]^{3/2}\approx
\frac{10^{15}}{\sin^3(r/2)}
\eqno(57a)
$$ 
Considering (56) with $\xi=2$ for $k=1$, the conditions for a good
approximation of (24) become 
$$\varepsilon_0^2\sin^2(r/2)\left(\frac{\rho}{\rho_0}\right)^{4/3}\approx10^{-20}
\left(\frac{\rho}{\rho_0}\right)^{4/3}\ll 1,\qquad  \Rightarrow\qquad
10^{-15}\frac{\rho}{\rho_0}\ll 1\eqno(57b)
$$
Thus, comparing (57a) and (57b), we see that the approximation to (24) is a good
one as long as one excludes very early evolution stages close to the FD density.
However, from (29b) and (31c), a ratio $\rho/\rho_0\approx 10^{15}$ can be
associated with to $T/\hat T_0\approx 10^{10}$, corresponding to
relativistic temperatures far above the range validity of the classical equation
of state (15), (24). Hence, the condition (57b) is perfectly reasonable and
allows for a large and physicaly justified range of validity for the
approximation of (27) to (24), based on the state variables (31). A convenient
limit of this validity range can be set for $\rho/\rho_0<10^6$, corresponding
approximately to $T/\hat T_0 <10^4$, roughly (and slightly larger than) the
range of post-decoupling temperatures for which the evolution a classical ideal
gas without interacting with radiation can be justified. The reasoning applied
above for the case $k=1$ is valid  for $k=0,-1$, as long as we consider only
spacetime sections with $r<2$. 

The range of validity of the ideal gas equation of state affects also the
question, raised in the previous section, on whether local observations on the
models are significantly different from those performed in dust FLRW
cosmologies. As far as the parameters $H_0,\,\Omega_0,\,q_0$ are concerned, the
small value $\varepsilon_0\approx 10^{-10}$ is sufficient to make them (via
(49), (50b)) practicaly identical to the equivalent FLRW parameters.
However, the red shift vs luminosity distance (54) contains products of
the form $\varepsilon_0F_0^2$ and $\varepsilon_0f_0^2$, and so for the
cases $k=0,-1$, these terms can be large even for such a small
$\varepsilon_0$. Hence, the same reasoning applies: for observations to
be close (up to order $\varepsilon_0$) to FLRW spacetimes, either we assume $k=1$
or, if $k=0,-1$, then we must only consider comoving observers near
$r=0$.         

The dynamical evolution of the models, especificaly the
model $k=1$, follows from the integral of the Friedmann equation (32), leading
to

$$H_0(t-t_i)={{1} \over {\sqrt{\nu_0}}}\int\limits_{A_i}^A  {{{\bar Ad\bar A}
\over {\left[ {\varepsilon _0/2+\mu _0\bar A-\bar A^2} \right]^{1/2}}}}={{1}
\over {\sqrt
{\nu _0}}}\left[ {\mu _0\arcsin \left( {{{2\bar A-\mu _0} \over {\sqrt {\mu
_0^2+2\varepsilon _0}}}} \right)-\left[ {\varepsilon _0/2+\mu _0\bar A-\bar A^2}
\right]^{1/2}} \right]_{A_i}^A\eqno(58a)$$
where
$$A\equiv \frac{R}{R_0}\qquad\nu _0={{c^2} \over
{(R_0H_0)^2}}={{2(\Omega _0-1)} \over  {2-\varepsilon _0}}={{2(2q_0-1)}
\over {4-\varepsilon
_0}},\quad \quad
\mu _0={{8\pi G}
\over {3c^4}}\rho _0R_0^2={{(2-\varepsilon _0)\Omega _0} \over {2(\Omega
_0-1)}}={{(2-\varepsilon _0)q_0+\varepsilon _0} \over {2q_0-1}}\eqno(58b) 
$$
and $A_i$ is the value of $A$ associated with a suitable initial surface
marked by $t=t_i$, for example the surface characterized by $T/\hat
T_0=10^4$ corresponding (from (31c)) approximately to $R/R_0\approx 10^{-2}$.
The regularity domain, starting (say) at $A_i=R_i/R_0= 10^{-2}$, must exclude
unphysical features like the FD singularity and ADS boundary,
marked by the coordinate values

$$\left[\frac{R}{R_0}\right]_{_{\rm{FD}}}=A_{_{\rm{FD}}}=
\sqrt{\frac{3\varepsilon_0}{2}}\sin(r/2)\le
\sqrt{\frac{3\varepsilon_0}{2}},\qquad
\hbox{FD singularity}\eqno(59a)
$$
$$\left[\frac{R}{R_0}\right]_{_{\rm{ADS}}}=A_{_{\rm{ADS}}}=
\sqrt{\frac{\varepsilon_0}{2}}\sin(r/2)\le
\sqrt{\frac{\varepsilon_0}{2}},\qquad
\hbox{ADS boundary}\eqno(59b)
$$

\noindent corresponding to very early stages of the evolution ($R/R_0\approx
10^{-5}$), well outside the stage where the initial value surface $R/R_0=
10^{-2}$ has been suggested. This surface is a regular Cauchy surface and the
evolution range of the models is given by 

$$A_{_{\rm{FD}}}\ll A_i=10^{-2} \le A \le A_{_{\rm{MAX}}}=\frac{\mu_0}{2}+
\sqrt{\left(\frac{\mu_0}{2}\right)^2+
\frac{\varepsilon_0}{2}}\approx \mu_0\eqno(60) 
$$
where $A_{_{\rm{MAX}}}$ marks the surface of ``maximal expansion'' characterized
by $H=\Theta/3=0$, after which the fluid recollapses following a time symmetric
pattern. From (58b) and (60), and bearing in mind that $\varepsilon_0\approx
10^{-10}$, we have $A_{_{\rm{MAX}}}\approx \Omega_0/(\Omega_0-1)$, a very close
fit to the value of $A_{_{\rm{MAX}}}$ obtained for a FLRW dust universe with
$k=1$.

The time required for the fluid to expand up to $A=1$ (the ``age of
the universe today'') and up to $A_{_{\rm{MAX}}}$ follows directly from (58) and
can be expressed as a function of either $(\Omega_0,\varepsilon_0) $ or
$(q_0,\varepsilon_0) $. As shown by figure 1, the curve relating $H_0(t_0-t_i)$
obtained from (58a-b) vs $\Omega_0>1$ for $\varepsilon_0=10^{-10}$ is almost
identical with the curve relating these same quantities for a FLRW dust universe
with $k=1$. As shown in Figure 2, for any given $\Omega_0 >1$ the value of
$H_0(t_0-t_i)$  obtained from a dust FLRW universe with $k=1$ is slightly larger
than this same quantity obtained from (58). In fact, the ratio of the
difference of these values of $H_0(t_0-t_i)$ to the FLRW value is of the order
of $10^{-4}$ of the latter value. Therefore, the age of the universe today
predicted by this model is practicaly the same as that predicted by a dust FLRW
universe with $k=1$, with $H_0(t_0-t_i)\to 2/3$  as $\Omega_0\to 1$, thus
sharing the same ``age problem'' associated with standard cosmology. Depending
on the value of $H_0$, in the range  $50-70\, \hbox{km}\,(\hbox{mpc s})^{-1}$,
standard cosmology leads to a range of age values $t_0-t_i\approx 10^{9-12}$
years that barely meets accepted age values obtained from globular clusters and
from nuclear and geological estimates\cite{peacock}. The same result follows for
models with $k=-1$ (assuming $r\ll 1 $), the ``age of the universe'' is 
almost the same as that predicted by standard cosmology with $k=-1$. The case
$k=0$ (also assuming $r\ll 1$) is slightly different, since in the FLRW models
$k=0$ necessarily implies $\Omega_0=1$, but for the $k=0$ Stephani case the
value $k=0$ corresponds to $\Omega_0<1$. We can only have $\Omega_0=1$ in the
Stephani models if $k=1$ and $\varepsilon_0=3(\gamma-1)$, a possible but
undesirable combination of values since $\varepsilon_0\approx 10^{-10}$ while
$\gamma=4/3,\,5/3$.    

From equations (31) for $k=1$ we have $p(t,0)=T(t,0)=0$ along the symmetry
center $r=0$, while both $p$ and $T$ have their maximal value along any surface
$t=\hbox{const}$ along the other symmetry center $r=\pi$ (see subsection C for
discussion on these points), that is, there is  a ``cold center'' along $r=0$ and
a ``hot center'' along $r=\pi$. However, if (28) and (56) hold and for matter
dominated conditions, the variation of $p$ and $T$ at any surface
$t=\hbox{const}$ is minimal (of the order $\varepsilon_0\approx 10^{-10}$) as
illustrated by the approximated forms of state variables at $t=t_0$  

$$\rho_0=\frac{mc^2N_0}{R_0^3},\quad p_0\approx \frac{N_0k_B\hat
T_0}{R_0^3}\sin^2(r/2)\left[1+O(\varepsilon_0)\right],\quad n_0\approx
\frac{N_0}{R_0^3}\left[1+O(\varepsilon_0)\right],\quad T_0\approx \hat
T_0\sin^2(r/2)\left[1+O(\varepsilon_0)\right]\eqno(61)
$$
clearly indicating a matter dominated regime $p_0/\rho_0\approx \varepsilon_0 \ll
1$, with very low pressure and temperature values providing an
excelent approximation to dust-like conditions. The models with
$\gamma=5/3$ might not be realistic becouse they do not allow  for the
description of another gas (CBMW radiation or other relique gases) as parts of a
decoupled mixture accompanying the monatomic ideal gas. However, if the matter
content of the present universe is dominated overwhelmingly by non-barionic CDM,
and if we can justify this CDM to satisfy the equation of state  of a classical
monatomic ideal gas ((15) and (24)), then the case
$\gamma=5/3$ could provide a nice toy model of a CDM universe in its latter
stages approaching near dust conditions. Obviuosly, it would be necessary in 
this case to infere the mass of the particles forming the CDM ideal gas, so
that an appropriate value of $\varepsilon_0$ be given.   

\subsection{The matter-radiation mixture.}

The state variables for the case $\gamma=4/3$ follow from (41a-c). This case is
perhaps more interesting than the case examined before because of the
possibility of accomodating a joint description of non-relativistic matter and a
relativistic relic gas. Considering the non-relativistic and ultra-relativistic
particles to be barions and photons, the ratio $N_0^{\rm{II}}/N_0^{\rm{I}}$ can
be associated with the ratio of photons to baryons, and so we have
$N_0^{\rm{II}}/N_0^{\rm{I}}\approx 10^8$ (considering $h\approx .6$ and
$\Omega_B\approx \Omega_0$\cite{kotu}). Assuming $m$ to be a proton mass and
$\hat T_0\approx 3$ degrees K., a good numerical estimate in (40) is
$\varepsilon_0\approx 3\times 10^{-5}$, a value five orders of magnitude larger
than
$\varepsilon_0$ in (30). However $\varepsilon_0$ is still sufficiently small so
that (38) might hold and (37) might provide a good approximation for (34) over a
wide temperature and density ranges. Since $T$ under (34a-b) is the  temperature
of the photon gas (so that $\hat T_0\approx 3$ degrees K), we can set again
$\rho_0$ to be matter-energy density today (overwhelmingly rest mass density).

The arguments discussed in the previous subsection regarding the convergence
of the  series in (27) for cases $k=0,-1$ apply also for the series in (37).
Hence, we consider only the case $k=1$, though the results can be applied to the
cases $k=0,-1$ for comoving observers near $r=0$. As before, we use look at
(46), now with the value $\varepsilon_0\approx 10^{-5}$, obtaining the allowed
values of $\rho/\rho_0$ (with the maximal value at the FD density) given by

$$\frac{\rho}{\rho_0}<\left[\frac{1}{2\varepsilon_0F^2}\right]^3\approx
\frac{10^{15}}{\sin^6(r/2)}
\eqno(62a)
$$ 
Considering (38) with $\xi=2$ for $k=1$, the conditions for a good
approximation to (34) become 
$$\varepsilon_0^2\sin^2(r/2)\left(\frac{\rho}{\rho_0}\right)^{2/3}\approx10^{-10}
\left(\frac{\rho}{\rho_0}\right)^{2/3}\ll 1,\qquad  \Rightarrow\qquad
10^{-15}\frac{\rho}{\rho_0}\ll 1\eqno(62b)
$$
Hence, from (62a) and (62b), a good approximation of (37) to (34a) requires
the exclusion of evolution stages close to the FD density.
From (39b) and (41c), the ratio $\rho/\rho_0\approx 10^{15}$ corresponds to
$T/\hat T_0\approx 10^{5}$, a higher value than temperatures of
the matter-radiation decoupling era, thus the range of
validity for the approximation of (37) to (34) exceeds that of a decoupled
mixture.

The dynamical evolution for the case $k=1$ follows from the following integral of
the Friedmann equation (42)

$$H_0(t-t_i)=\sqrt {1-\varepsilon _0}\int\limits_{A_i}^A  {{{\sqrt {\bar A}d\bar
A} \over {\sqrt {\Omega _0-\varepsilon _0-(\Omega _0-1)\bar A}}}}=\sqrt
{{{1-\varepsilon _0} \over {\Omega _0-1}}}\left[ {\xi _0\arcsin \left( {{{\bar
A} \over {\xi _0}}-1} \right)-\bar A^{1/2}\left( {2\xi _0-\bar A} \right)^{1/2}}
\right]_{\bar A=A_i}^{\bar A=A}\eqno(63)
$$
$$\hbox{where:}\qquad A\equiv \frac{R}{R_0},\qquad \xi _0\equiv {{\Omega
_0-\varepsilon _0}
\over {2(\Omega _0-1)}}
$$
The regularity domain, starting at $\rho/\rho_0\approx 10^{15}$ or
$R/R_0\approx10^{-5}\approx\varepsilon_0$ must exclude the FD singularity and ADS
boundary  marked by 
$$
\left[\frac{R}{R_0}\right]_{_{\rm{FD}}}=A_{_{\rm{FD}}}=
2\varepsilon_0\sin^2(r/2)\le
2\varepsilon_0,\qquad
\hbox{FD singularity}\eqno(64a)
$$
$$\left[\frac{R}{R_0}\right]_{_{\rm{ADS}}}=A_{_{\rm{ADS}}}=
\varepsilon_0\sin^2(r/2)\le
\varepsilon_0,\qquad
\hbox{ADS boundary}\eqno(64b)
$$
indicating that we can set $A_i=A(t_i)$ to a value greater than
$2\varepsilon_0\approx 10^{-5}$ that would correspond to the decoupling era ($
T/\hat T_0\approx 10^3$), for example: $A_i=10^{-3}$. The regular evolution
range of the model extends from $A_i$ up to the value $A_{_{\rm{MAX}}}$, marking
the maximal expansion $\Theta=0$, where the fluid bounces and begins to collapse
in a time symmetric pattern. Evaluating
$A_{_{\rm{MAX}}}$ from (63) yields

$$A_{_{\rm{MAX}}}=2\xi_0={{\Omega
_0-\varepsilon _0}
\over {\Omega _0-1}} \approx \frac{\Omega_0}{\Omega_0-1}
\eqno(65)$$  
a value almost equal to that obtained for a dust FLRW universe with $k=1$.
The ``age of the universe today'' can be evaluated by setting $A=1$ in (63).
Ploting the resulting value $H_0(t_0-t_i)$ vs $\Omega_0>1$ and comparing with
the FLRW spacetime leads to curves almost identical to those shown in figure 1.
However, the ratio of differences of $H_0(t_0-t_i)$ to the FLRW value
yields an even closer fit to the FLRW expression (see figure 2). Hence, as
in the case $\gamma=5/3$, this model presents the same ``age problem'' of
standard cosmology mentioned before. If we assume that the non-relativistic
component  includes cold dark matter (assuming also that it can be described as
an ideal gas), we would have a value
$N_0^{\rm{I}}$ larger by two orders of magnitude, modifying (40) to
$\varepsilon_0\approx 10^{-7}$. However, this alternative value of
$\varepsilon_0$ does not change the close fit to the dust FLRW universe, in
fact, it makes this fit even closer, since the only way to obtain values of
$H_0(t_0-t_i)$ significantly different from the dust FLRW values is to have
a larger $\varepsilon_0\approx 1$, though the resulting values of
$H_0(t_0-t_i)$ are then {\it significantly smaller} than those of the FLRW
model. As in the case $\gamma=5/3$, this hopeless situation also occurs
for the cases with $k=0,-1$.

As in the previous case, from (39) and (41), we also have a ``cold'' symmetry
center $r=0$ where $p(t,0)=T(t,0)=0$ (see subsection C for discussion on this
aspect), and a ``hot'' center at $r=\pi$ where both $p$ and $T$ have their
maximal value along any surface $t=\hbox{const}$. At $t=t_0$ the state variables
take the approximated forms

$$\rho_0=\frac{mc^2N_0^{\rm{I}}}{R_0^3},\quad p_0\approx
\frac{N_0^{\rm{II}}k_B\hat
T_0}{R_0^3}\sin^2(r/2)\left[1+O(\varepsilon_0)\right],\quad n_0\approx
\frac{(N_0^{\rm{I}}+N_0^{\rm{II}})}{R_0^3}\left[1+O(\varepsilon_0)\right],\quad
T_0\approx
\hat T_0\sin^2(r/2)\left[1+O(\varepsilon_0)\right]\eqno(66)
$$
so that matter dominated dust-like conditions do prevail at $t=t_0$, with the
variation of $p$ and $T$ of the order $\varepsilon_0\approx 10^{-5}$. Also, as
with the case $\gamma=5/3$, as long as $F_0$ and $f_0$ in (54) are bonded,
observation parameters are almost identical to their FLRW equivalents. However,
the pressure and temperature in this case correspond to the pressure and
temperature of the MWCB (with $\hat T_0=3$ degrees K.). Hence, the ratio
$p_0/\rho_0\approx \varepsilon_0\approx 10^{-5}$ from (66), is roughly the
accepted value of ratio of the energy density of the CMBR ($\approx 3p_0$) to
that of visible non-relativistic matter. Regarding the temperature, a deviation
of order $\varepsilon_0\approx 10^{-5}$ gives an excellent fit with the observed
deviations of the isotropy of the temperature of the MWCB\cite{peacock},
\cite{kotu}. We might consider including CDM in the matter content of the
non-relativistic component so that $\varepsilon_0\approx 10^{-7}$.
This might sound reasonable, but CDM is most possibly non-barionic, and so, as
in the case $\gamma=5/3$, we would have to justify that it satisfies an equation
of state like (34). With or without CDM, the case $\gamma=4/3$ could be a
nice toy model of a matter dominated universe approaching near dust
conditions, though it would be a better model than that provided by $\gamma=5/3$
as it would include the MWCB.

\subsection{The drawbacks.}
 It emerges from the discussion in the previous subsections that the $k=1$
models, whether $\gamma=4/3$ or $\gamma=5/3$, share two unappealing features: 
(A) The maximal and minimal values of the rest mass density ($mc^2n$ or
$mc^2n^{\rm{I}}$) respectively occur along the symmetry centers $r=0$ and
$r=\pi$, while the maximal and minimal values of the internal energy density
($\rho-mc^2n\approx 3p/2$ or $\rho-mc^2n^{\rm{I}}\approx 3p$) occur along $r=\pi$
and
$r=0$.  This means that along $r=0$ all energy density is rest mass energy
density and there is a higher density of particles but zero temperature and
pressure for all times (see discussion further ahead), while at $r=\pi$ there
are less particles by volume but a higher temperature and pressure. (B) The fact
that pressure and temperature vanish along $r=0$ for all evolution times (forced
by the specific spacial dependence of the metric contained in the function $F$)
is a very different situation from $p,\,T$ tending to zero asymptoticaly as the
fluid expands and cools. This situation leads to the entropy per particle (18c)
diverging as $r\to 0$, so that the left hand side of the Gibbs equation (14) 
($T s'$), remains finite (as the right hand side does). Although this feature is
problematic, it is consistent with the fact that an absolute zero temperature
(if $T$ is stricly zero instead of $T\to 0$ asymptoticaly) is impossible to
attain in equilibrium classical thermodynamics\cite{huang}. This 
feature (appearing also for the cases $k=0,-1$) is perhaps the strongest 
drawback on the physical plausibility of the models.

\subsection{Final remarks.}
The models derived and presented in this paper are mathematicaly
simple, theoreticaly consistent and satisfy physicaly motivated equations of
state. Whether one considers it a virtue or a defect, the models bear an
extremely close resemblance (observationaly and dynamicaly) to FLRW
cosmologies in the matter dominated regime with a dust source, therefore they
 share the possible failure of matter dominated standard cosmology in
dealing with recent observational evidence\cite{peacock}. However, even if hard
core cosmologists might dismiss them as not sufficiently realistic, the
solutions can at least be applied as self-consistent cosmological toy models or
as useful educational tools in teaching cosmology. Hopefuly the results of the
present paper, besides rescueing the Stephani Universes from oblivion, will
motivate and improve further research based on this nice and elegant class of
solutions.              

\acknowledgements
This work is supported by grant DGAPA-IN122498. The wise meowing and feline
wisdom of Mimi, Tontina, Ni\~na, Chata, Pintita, Martin, Feucho, Bonito, Ulises,
Vitorio and Pocholo provided the author with enough inspiration as to make the
writing of this paper an enjoyable venture.

\begin{figure}
\caption{{\bf Age of universe today.}
This figure displays $H_0(t_0-t_i)$ vs $\Omega_0$ for the $k=1$
model with $\gamma=5/3$ (solid line) and for a $k=1$ FLRW cosmology with a
dust source. The values of $H_0(t_0-t_i)$ follow from (58) with $A=1$ and
$A_i=10^{-3}$, with $H_0=60\,\hbox{km (mpc seg)}^{-1}$. Notice the almost
identical fit between both cosmological models. The same close fit (not
displayed) with standard cosmology occurs for
$k=-1$, as well as for the $k=1,\,k=-1$ models with $\gamma=4/3$. For the
case $k=0$ standard cosmolgy yields $\Omega_0=1$, while for the Stephani
models (either $\gamma=4/3,\,5/3$) we would obtain a plot analogous to the case
$k=1$ with $\Omega_0>1$. These (and the remaining) plots were obtained with the
help of the symbolic computing package Maple V.}\label{f1}
\end{figure}

\begin{figure}
\caption{{\bf The relative ratio of the differences of $H_0(t_0-t_i)$ for the
Stephani and FLRW models.} This plot displays the ratio
$X_0\equiv\log_{10}|\tilde Y_0-Y_0|/\tilde Y_0$ as a function of $\Omega_0$ for
the case $k=1$ of $\gamma=4/3,\,5/3$, where $Y_0\equiv H_0(t_0-t_i)$ and $\tilde
Y_0$ is the FLRW equivalent. The smallness of this ratio illustrates the close
fit in the dynamics of the Stephani models and standard cosmology. The
resemblance is tighter for  $\gamma=4/3$ (curve below) than for the model
$\gamma=5/3$ (curve above). }\label{f2}
\end{figure}

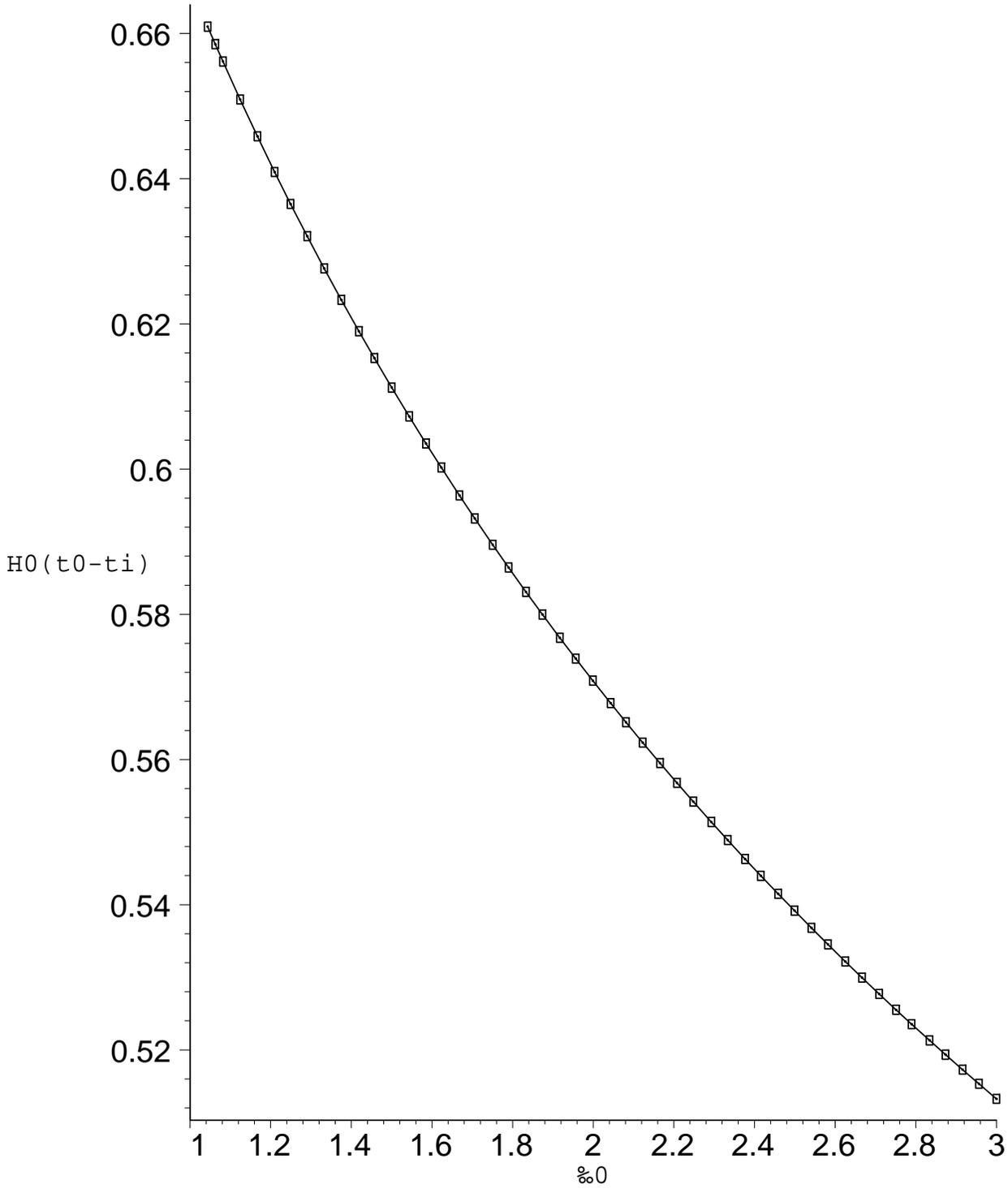
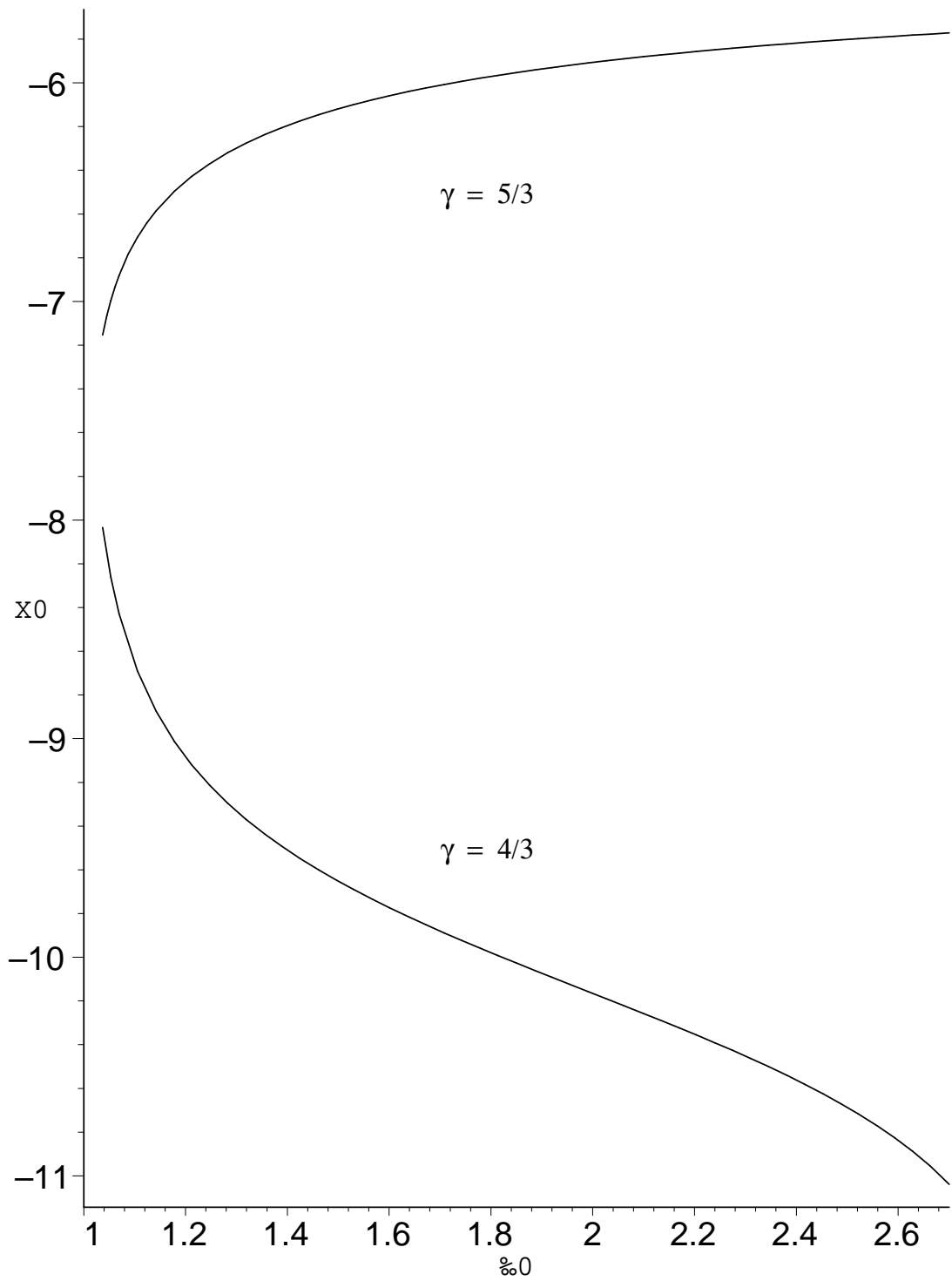
\end{document}